\documentclass[aps,pra,showpacs,twocolumn]{revtex4}
\usepackage{amssymb}
\usepackage{epsfig}
\usepackage{graphicx}
\usepackage{amsmath}
\usepackage{array,color}
\usepackage{mathrsfs}
\usepackage{subfigure}
\newenvironment{sequation}{\small\begin{equation}}{\end{equation}}

\begin{document}
\title{Spin-orbit coupling induced separation and hidden spin textures in spin-$1$ Bose-Einstein condensates}
\author{Shu-Wei Song$^1$, Yi-Cai Zhang$^1$, Qing Sun$^1$, Hanquan Wang$^2$, Lin Wen$^1$, A. C. Ji$^{3,1}$ and W. M. Liu$^{1}$}
\address{$^1$Beijing National Laboratory for Condensed Matter Physics,
Institute of Physics, Chinese Academy of Sciences, Beijing 100190,
China}
\address{$^2$School of Statistics and Mathematics, Yunnan University of Finance and Economics,
Kunming, Yunnan Province, 650221, China}
\address{$^3$Department of Physics, Capital Normal University, Beijing 100048, China}
\date{\today}
\begin{abstract}
We analytically and numerically investigate the ground state of the spin-orbit coupled spin-$1$ Bose-Einstein condensates in an external parabolic potential. When the spin-orbit coupling strength $\kappa$ is comparable with that of the trapping potential, the density distribution centers of different components of the spinor condensate deviate evidently from the trap center in the plane wave and stripe phases. When $\kappa\gg1$, the magnitude of this deviation decreases as $\kappa$ is getting larger and larger. Correspondingly, periphery half-skyrmions textures arise. This deviation can be reflected by the non-uniform magnetic moment in the $z$ direction, $\mathcal{F}_z$. With the manipulation of the external trap, the local magnitude of $\mathcal{F}_z$ can be increased evidently. This kind of increase of $\mathcal{F}_z$ is also observed in the square vortex lattice phase of the condensate.
\end{abstract}
\maketitle
\section{INTRODUCTION}\label{sec1}
The spin-orbit (SO) coupling is an essential mechanism for most spintronics devices, and leads to many fundamental phenomena in condensed matter physics and atomic physics. For example, SO coupling gives rise to quantum spin Hall effect in electronic condensed matter systems \cite{Shuichi}. Recently, artificial external Abelian or non-Abelian gauge potentials coupled to neutral atoms have been generated by controlling atom-light interaction \cite{LinPRL, LinNature, Aidelsburger,DLCampbell}, which provides the possibility of spin control in neutral atom systems through laser fields. Without the SO coupling, spinor Bose-Einstein condensates (BECs) have received extensive explorations, including the spin dynamics and textures \cite{SMuhlbauer,HMPrice,Leanhardt,Wright,Stenger,UAKhawaja,LESadler,Chang,LLi,Beata,Wenxian,AnChun,Jae-yoon,JGuzman,shuwei}. In the presence of the SO coupling, the neutral atoms in the effective abelian and non-abelian gauge fields behave like electrons in an electro magnetic field or electrons and abundant phenomena arise. Thus, SO coupled cold atom systems have attracted intensive attention since last few years \cite{Tudor,Sarang,Subhasis,HuiPRL,JRadic,HuiHu,Dan-Wei,BRamachandhran,Ryan,Wei,Chao-Ming,Yan-Xiong,MMerkl,Yongping,SWSu,Deng,Xiao-Qiang,Xiang-Fa,ZFXu}.
For example, the condensate may become self-trapped as a result of the spin-orbit coupling and the nonlinearity, resembling the so-called chiral confinement \cite{MMerkl}. Dynamical oscillations of the condensate with SO coupling are studied in \cite{Yongping}, where oscillation period, similar to the Zitterbewegung oscillation, was found. Besides, SO coupled BECs with dipole-dipole interactions \cite{Deng} and rotating trap \cite{Xiao-Qiang,Xiang-Fa}, have also been studied. In \cite{ZFXu}, SO coupled atomic spin-$2$ BEC was investigated, and square or triangular density patterns were reported. Both patterns evolve continuously into striped forms with increased asymmetry of the SO coupling.

The SO coupled spinor condensate develops a spontaneous plane wave phase or stripe phase because of the interaction energy \cite{Ho2011,Wang2011,SKYip}. In this case, the harmonic length of the external trap is much greater than the wavelength of the stripe, indicating a weak external trap compared to the SO coupling strength. In two components condensate, the phase diagram including half-quantum vortex, vortex lattice and stripe states has been obtained in Ref. \cite{Subhasis}. For the case with strong external traps and strong spin-orbit coupling, half-quantum vortex state and Skyrmion lattice patterns can arise \cite{HuiPRL}, where the atomic interaction is relatively weak. However, spin-$1$ BECs in the presence of external trap with neither too strong nor too weak SO coupling strength, i.e., $\kappa\sim1$, are less studied. The atom-atom interaction needs to be large enough to produce trapped plane wave or stripe state in an external trap. In the trapped plane wave or stripe state, there exist nontrivial detailed structures, as we will see in the following sections.

In the present paper, we both analytically and numerically investigate certain characteristics of the spin-$1$ BEC with SO coupling strength $\kappa\sim1$ in external parabolic potential. Comparing with the homogenous case, we find that the ground state wave functions of different components of the condensate dislocate and exhibit finite separation, which is reflected by the non-uniform magnetic moment distribution in $z$ direction. In particular, we find associated hidden spin textures in the usual plane wave and stripe phases. Also, effects of the manipulation
of external potentials on the magnetic moment in $z$ direction, are explored in the trapped plane wave, stripe and the square
lattice structure phases. The separation in the trapped stripe state is more complex. By applying an anisotropic external potential, the trapped stripe state develops a spin texture characterized by aligned half-skyrmions.

The rest of the paper is organized as follows. In Sec. \ref{sec2}, we propose an variational approach to study the separation of density distributions between different components in the plane wave phase. In Sec. \ref{sec3}, we numerically confirm the analytical results obtained in Sec. \ref{sec2} and exhibit the hidden spin texture in the plane wave phase.
Sec. \ref{sec4} focuses on the hidden spin textures in stripe phase. Finally, the conclusions are summarized in Sec. \ref{sec5}.

\section{theoretical framework}\label{sec2}
We consider the quasi-$2$ dimensional spin-$1$ BECs with Rashba SO coupling in the $x-y$ plane confined in an external parabolic potential. Without regarding to the interaction, the single particle part of the Hamiltonian without external
potential is
\begin{eqnarray}
\hat{H}_0& = & \int d\mathbf{r}\left[\mathbf{\Psi}^\dag\left(\frac{\hat{\mathbf{p}}^2+2\kappa\hat{\mathbf{p}}\cdot\hat{\mathbf{F}}}{2M}    \right)\mathbf{\Psi}\right], \label{eq1}
\end{eqnarray}
where $\hat{\mathbf{F}}=(F_x,F_y)$ is spin-1 representation of Pauli matrices.
%, with
%\begin{gather*}
%F_x=\begin{pmatrix} 0 & 1 & 0\\ 1 & 0 & 1\\ 0 & 1 & 0 \end{pmatrix}\quad
%F_y=\begin{pmatrix} 0 & -i & 0\\ i & 0 & -i\\ 0 & i & 0 \end{pmatrix}\quad.
%\end{gather*}
The SO coupling strength $\kappa$ is experimentally related to the wavelength of the laser beams. The three components spinor is $\mathbf{\Psi}=(\psi_1,\psi_0,\psi_{-1})^T$ (the superscript $T$ stands for the transpose) and $M$ is the atomic mass.
\begin{figure}[tbp]
\includegraphics[width=7cm]{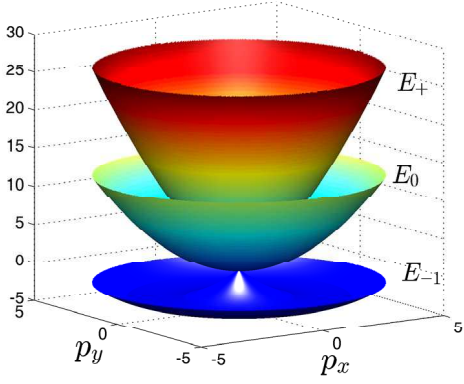}
\caption{Schematic picture of the band structure with SO coupled cold atoms.}
\label{fig1}
\end{figure}

In order to gain some intuition, we first focus on the homogeneous condensate. Thus, the single-particle spectrum of the Hamiltonian (\ref{eq1}) has the following three branches with different helicity in the momentum space (see Fig. \ref{fig1}).
\begin{eqnarray}
E_0&=&\frac{\mathbf{p}^2}{2M}\notag,\\
E_{\pm1}&=&\frac{1}{2M}\left(\mathbf{p}^2\pm 2\sqrt{2}\kappa|\mathbf{p}|\right). \notag
\end{eqnarray}
The single-particle ground state is in the negative helicity branch with $p_0=|\mathbf{p}|=\sqrt{2}\kappa$, and the wave function is given by
\begin{equation}\renewcommand\arraystretch{1.5}
u_{-}=\left(
\begin{array}{c}
 -\frac{1}{2}e^{-i\phi_{\mathbf{p}}}\\
\frac{1}{\sqrt{2}}\\
-\frac{1}{2}e^{i\phi_{\mathbf{p}}}
\end{array}
\right) \notag,
\end{equation}
where $\phi_{\mathbf{p}}= \arg(p_x+ip_y)$. All the states with the same $|\mathbf{p}|$ are circularly degenerate.

Taking account of the interaction and the external potential, the mean-field version of the total Hamiltonian is
\begin{equation}\label{eq2}
\begin{split}
H=&\int d\mathbf{r}\bigg\{\sum_{m=0,\pm1}\big[-\frac{1}{2}\psi_m^*\nabla^2\psi_m+\frac{1}{2}(x^2+y^2)|{\psi_m}|^2\big]\\
&+\kappa\big[\psi_1^*(-i\partial_x-\partial_y)\psi_0+\psi_0^*(-i\partial_x-\partial_y)\psi_{-1}\\
&+\psi_0^*(-i\partial_x+\partial_y)\psi_1+\psi_{-1}^*(-i\partial_x+\partial_y)\psi_0\big]\\
&+\frac{c_0}{2}(|\psi_1|^2+|\psi_0|^2+|\psi_{-1}|^2)^2\\
&+\frac{c_2}{2}\big[(|\psi_1|^2-|\psi_{-1}|^2)^2+2|\psi_1^*\psi_0+\psi_0^*\psi_{-1}|^2\big]\bigg\},
\end{split}
\end{equation}
where the coupling constants $c_0$ and $c_2$
characterize the density-density and spin-exchange interactions, respectively. The length, time and energy are respectively scaled in units of $\sqrt{\hbar/M\omega}$, $1/\omega$ and $\hbar\omega$. $\omega$ is the radial trapping frequency.
\begin{figure}[tbp]
\includegraphics[width=9cm]{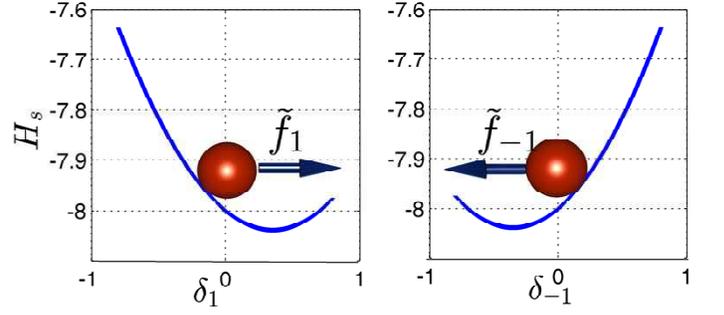}
\caption{(Color online) The SO-coupling-energy function $H_s$ of (a) $\delta_1$ and (b) $\delta_{-1}$ with $\eta=0.0096$ and $\kappa=2.0$. The light-red balls at points $\delta_1=\delta_{-1}=0$ feel forces pointing to the equilibrium position.}
\label{fig2}
\end{figure}

To investigate this hidden separation of the plane wave phase in the direction perpendicular to the propagation of the plane wave, we suppose a plane wave propagating along $x$ axis and use the ansatz:
\begin{equation}\renewcommand\arraystretch{1.5}
\left(
\begin{array}{c}
\psi_1\\
\psi_0\\
\psi_{-1}
\end{array}
\right) =Ce^{ip_0 x}\left(
\begin{array}{c}
 -\frac{1}{2}e^{-p_0^2\eta(x^2+(y-\delta_{1})^2)}\\
\frac{1}{\sqrt{2}}e^{-p_0^2\xi(x^2+(y-\delta_{0})^2)}\\
-\frac{1}{2}e^{-p_0^2\eta(x^2+(y-\delta_{-1})^2)}
\end{array}
\right), \label{eq3}
\end{equation}
where $\eta$ and $\xi$ are positive dimensionless variational parameters related to the width of the condensate, and $C=2\,{\frac {p_0\,\sqrt {\eta\,\xi}}{\sqrt {\pi \, \left( \xi+\eta \right) }}}$ is the normalization coefficient. Here we introduce the deviation $\delta_m$ from the origin in the $y$ direction. A deviation in the $x$ direction can also be supposed and this deviation is found to be zero to minimize the total mean field energy.

Without the SO coupling term in the Hamiltonian, the condensate feels comfortable when the distribution centers of the three components occupy the center of the external parabolic potential. It is not the same situation any longer once the SO coupling is included. In the presence of the SO coupling, the gaussian-profile density distribution with $\delta_{1}=\delta_{0}=\delta_{-1}=0$ is not the ground state . We can introduce the effective virtual force to determine the deviating direction of gaussian-profile density of each component. Thus, the mean-field energy would be minimized. We define the effective virtual force: $\tilde{f}_{i}=-\partial H_{s}/\partial\delta_{i}|_{\delta_{i}=0}$ $(i=0,\pm1)$. After substituting the ansatz in Eq. (\ref{eq3}) into the SO part of the total energy, we have

\begin{figure}[tbp]
\includegraphics[width=9.5cm]{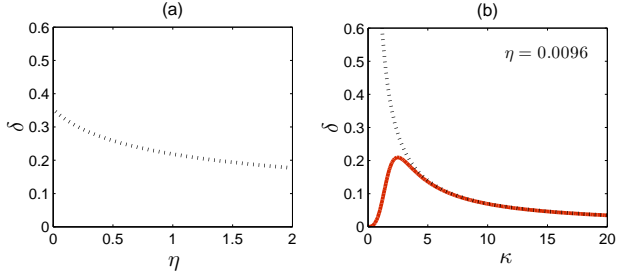}
\caption{(Color online) (a) Dependence of $\delta$ on $\eta$ with $\kappa=2.0$. (b) Dependence of $\delta$ on $\kappa$ with $\eta=0.0096$. The black dotted line corresponds to the case when only the SO part of the mean field energy is minimized. The solid red line corresponds to the case of counting in both the SO and the external potential parts of the mean field energy.}
\label{fig3}
\end{figure}

\begin{equation}\renewcommand\arraystretch{1.5}
\begin{split}
\tilde{f}_{1}=&8\sqrt {2}\,{\frac {{\kappa}^{3}{\eta}^{2}{\xi}^{2}}{ \left( \xi+\eta
 \right) ^{3}}},\\
\tilde{f}_{0}=&0,\\
\tilde{f}_{-1}=&-8\sqrt {2}\,{\frac {{\kappa}^{3}{\eta}^{2}{\xi}^{2}}{ \left( \xi+\eta
 \right) ^{3}}}.
\label{eq7}
\end{split}
\end{equation}
Since $\eta$, $\xi$ and $\kappa$ are positive parameters, the direction of the dragging force have been fixed. The component with magnetic quantum number $m=1$ feels a dragging force pointing to the left side of the propagating plane wave direction and the component with magnetic quantum number $m=-1$ feels the same force but in opposite direction, while no force acts on the components with magnetic quantum number $m=0$. Based on these effective virtual forces, we suppose $\delta_{1}=-\delta_{-1}=\delta$, $\delta_{0}=0$, and $\xi=\eta$. In fact, further calculations show that the deviation is not sensitive to the relative width of the condensate for the present concerned range of values of the SO coupling strength. After substituting the ansatz in Eq. (\ref{eq3}) into the SO part of Eq. (\ref{eq2}), we have

\begin{figure}
\centering
\subfigure{(a)}
\begin{minipage}[b]{0.5\textwidth}
\centering
\includegraphics[width=3.8in]{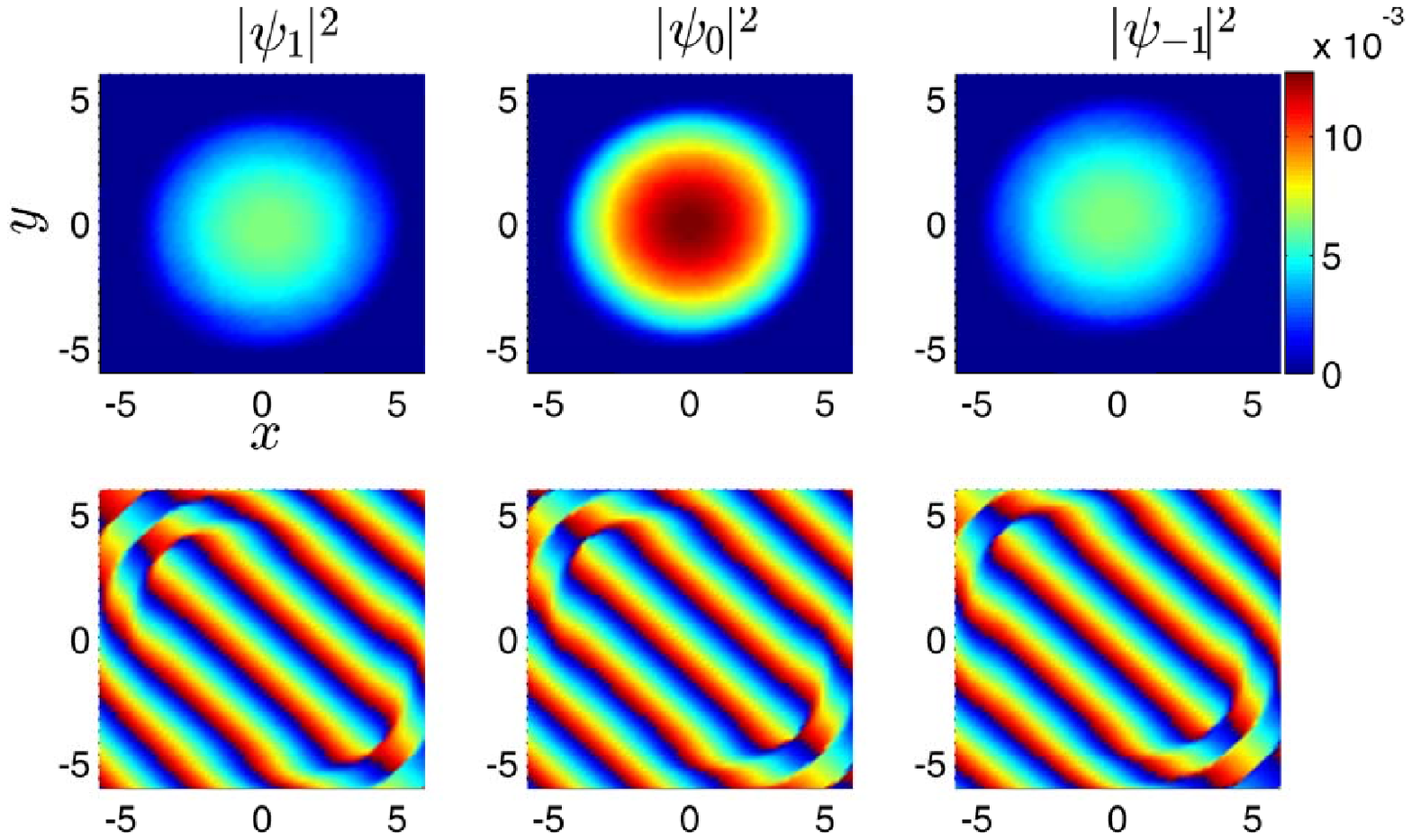}
\end{minipage}
\subfigure{(b)}
\begin{minipage}[b]{0.5\textwidth}
\centering
\includegraphics[width=2.0in]{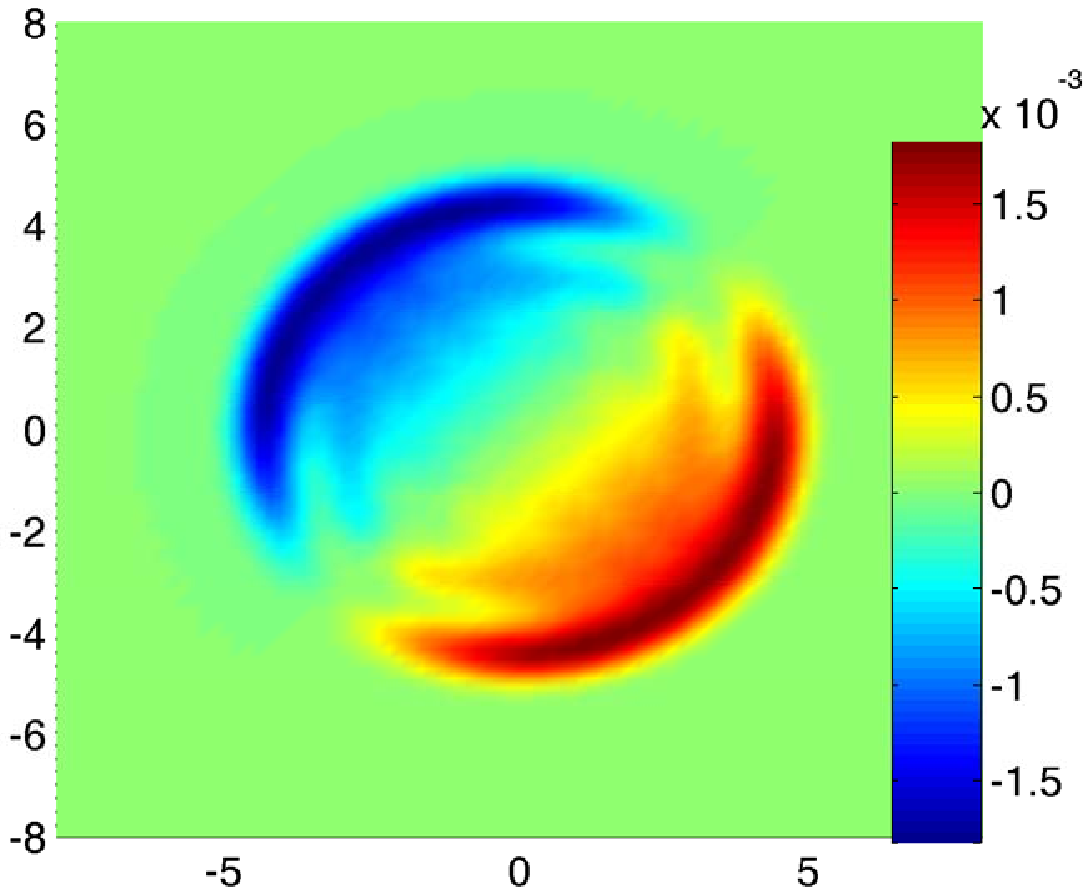}
\end{minipage}
\caption{(Color online) (a) The density (upper panels) and phase (bottom panels) distributions of the plane wave phase with $\kappa=2.0$, $c_0=500.0$ and $c_2=-5.0$. (b) The non-uniform density of magnetic moment along the $z$ direction, $\mathcal{F}_z$, in the plane wave phase.}
\label{fig4}
\end{figure}

\begin{eqnarray}
H_s=-\kappa\,{{\rm e}^{-1/2\,\eta\,{p_0}^{2}{\delta}^{2}}}\sqrt {2}
 \left( p_0+\eta\,{p_0}^{2}\delta \right) \label{eq4}.
\end{eqnarray}
Starting from $H_s$, we can find that the minimum functional energy requires the condition:
\begin{eqnarray}
\delta=\frac {-1+\sqrt {1+4\,\eta}}{2\sqrt{2}\eta\,
\kappa}>0
. \label{eq5}
\end{eqnarray}
In Fig. \ref{fig2}, the SO part of the total mean-field energy with respect to $\delta_{1}$ ($\delta_{1}=\delta$) and $\delta_{-1}$ ($\delta_{-1}=-\delta$) is plotted. The light-red balls, which deviate from the equilibrium position feel forces as expressed in Eq. (\ref{eq7}).
We know that spin currents accompany charge currents in high-mobility two-dimensional electron systems
\cite{SMurakami}, and thus spin accumulation will be induced with the presence of the SO coupling.
Starting from the single particle Hamiltonian in Eq. (\ref{eq1}), we can obtain the force felt by the components of the
condensate in the Heisenberg picture by following calculations in Ref. \cite{JohnSchliemann}:
\begin{equation}
\vec{f}=2\kappa^2/(\hbar M^2)\left(\hat{\mathbf{p}}\times \hat{e}_{z}\right){F}_{z}.\label{eq0}
\end{equation}
Here we should mention that, the above introduced effective virtual forces in Eq. (\ref{eq7}) come from small deviations from the ground state (equilibrium position), which is not the same as the true force of the particles during the dynamical motions, as expressed in Eq. (\ref{eq0}).

The dependence of $\delta$ on $\eta$ and $\kappa$ is plotted in Fig. \ref{fig3}. From Fig. \ref{fig3}(a) with $\kappa=2.0$, we can find
that the deviation does not change dramatically with respect to $\eta$ and chooses values $0.2\sim0.35$. More calculations show that the deviation function of SO coupling strength is $\delta=1/\sqrt{2}\kappa$ as $\eta\rightarrow0$ (meaning an unlimited condensate width corresponding to the case of zero external potential). This limit behavior is not affected by the interactions between particles.
Figure \ref{fig3}(b) shows the dependence of $\delta$ on $\kappa$ with $\eta=0.0096$. For the black dotted line, only the SO part of the mean field energy is minimized.  As $\kappa$ increases, i.e., is approaching the strong SO coupling regime, the deviation diminishes at a rate proportional to $1/\kappa$.
From Fig. \ref{fig3}(b), we can see from the black dotted line that the deviation diverges as the SO coupling strength getting weaker and weaker ( $\kappa\rightarrow0$), which is nonphysical. This divergence can be canceled by counting in the external potential, as shown by the solid red line of Fig. \ref{fig3}(b).
\begin{figure}
\centering
\subfigure{(a)}
\begin{minipage}[b]{0.5\textwidth}
\centering
\includegraphics[width=2.4in]{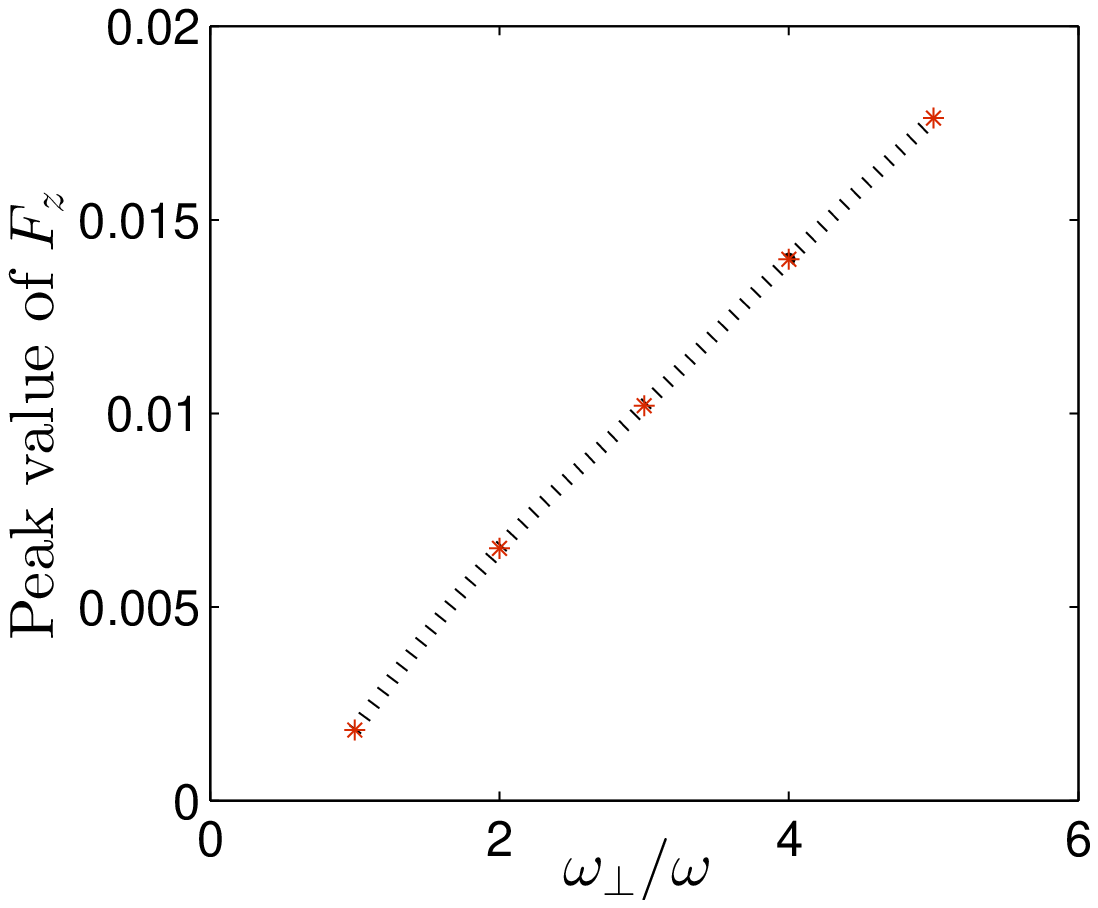}
\end{minipage}
\subfigure{(b)}
\begin{minipage}[b]{0.5\textwidth}
\centering
\includegraphics[width=2.2in]{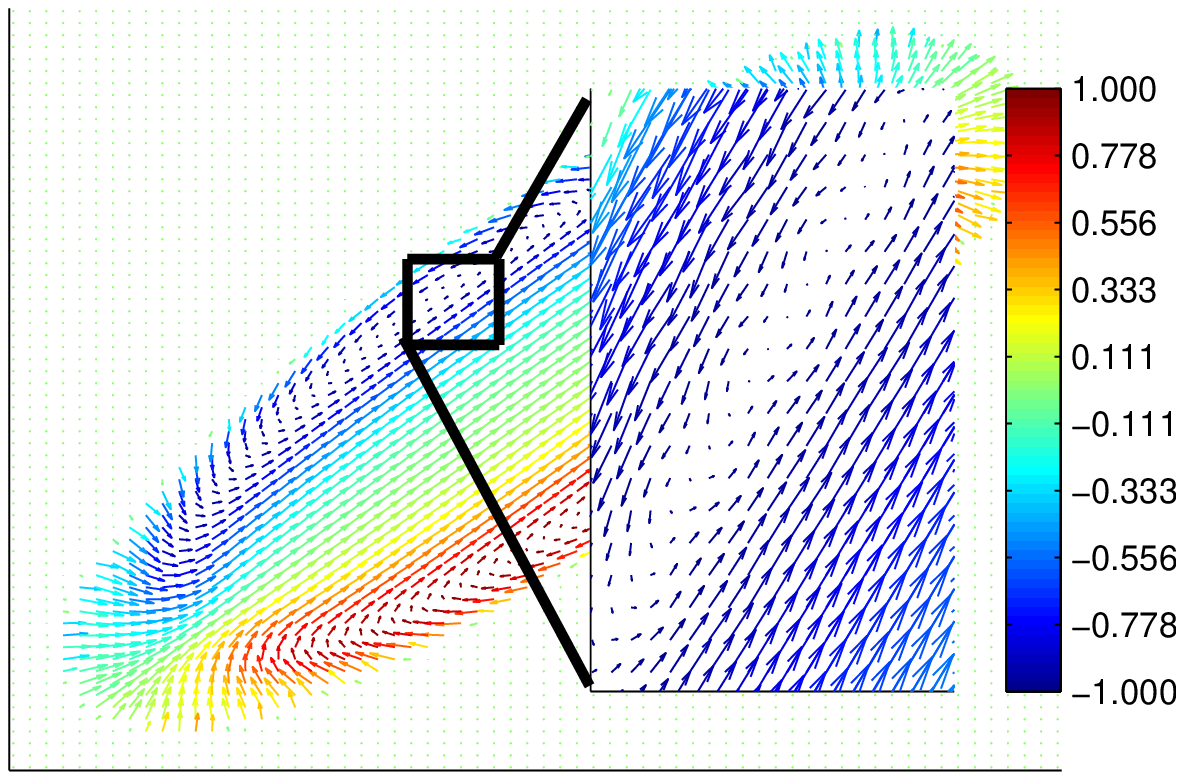}
\end{minipage}
\caption{(Color online) (a) As the condensate is being squeezed, the peak value of the $\mathcal{F}_z$ increases with respect to $\omega_\perp/\omega$, where $\omega_\perp$ is the trapping frequency in the squeezing direction. (b) The corresponding spin texture, showing the hidden half-skyrmions in the periphery of the condensate in the plane wave phase.  The relevant parameters are $\kappa=2.0$, $\omega_\perp/\omega=4.0$, $c_0=500.0$ and $c_2=5.0$.}
\label{fig5}
\end{figure}

\begin{figure}[tbh]
\includegraphics[width=7.80cm]{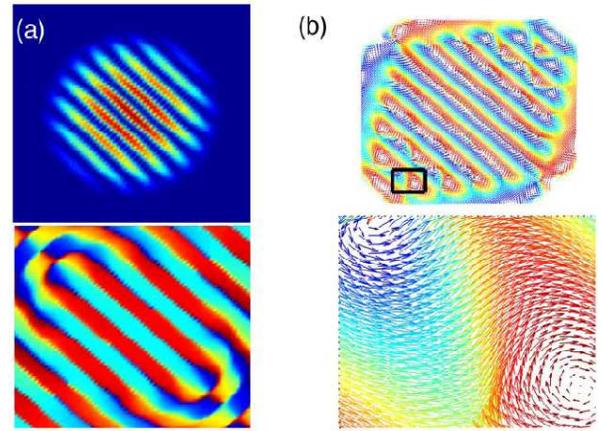}
\caption{(Color online) (a) The density and phase distributions in the stripe phase and (b) the corresponding spin texture, showing the hidden half-skyrmions in the periphery of the condensate with $\kappa=2.0$, $c_0=500.0$ and $c_2=5.0$. Spin texture in the bottom is the magnification of the squared part in the top}
\label{fig6}
\end{figure}
To minimize the mean-field energy, there needs to be deviation of density distribution between different components of the condensate, resulting in polarized magnetic momentum distribution ($\mathcal{F}_{z}$). It is necessary to notice that the polarized magnetic momentum distribution has nothing to do with spin-dependent force expressed in Eq. (\ref{eq0}). Through minimizing the mean-field energy, the polarized magnetic momentum distribution in the plane wave phase is the consequence of the cooperation of all the components of the condensate in the presence of the SO coupling and the external trap.

For the stripe phase of the condensate, the matter wave is composed of two coherent wave vector
states that propagating face to face along the $x$ axis. We can choose the ansatz:
\begin{equation}\renewcommand\arraystretch{1.5}
\left(
\begin{array}{c}
 \psi_1\\
\psi_0\\
\psi_{-1}
\end{array}
\right) =C\left(e^{ip_0 x}-e^{-ip_0 x}\right)\left(
\begin{array}{c}
 -\frac{1}{2}\,{{\rm e}^{-{p_0}^{2}\eta\, \left( {x}^{2}+ \left( y+\delta
 \right) ^{2} \right) }}\\
 \frac{\sqrt{2}}{2}{{\rm e}^{-{p_0}^{2}\eta\, \left( {x}^{2}+{y}^{2}
 \right) }}\\
-\frac{1}{2}\,{{\rm e}^{-{p_0}^{2}\eta\, \left( {x}^{2}+ \left( y-\delta
 \right) ^{2} \right) }}
\end{array}
\right), \label{eq8}
\end{equation}
where $C=p_0\,\sqrt {\frac{\eta}{\pi}}
$ is the normalization coefficient. Starting from the SO part of Eq. (\ref{eq2}), the deviation formula similar to Eq. (\ref{eq5}) can be found:
\begin{equation}
\sqrt {2}{\kappa}p_{0}^{3}\eta\,\delta\,{{\rm e}^{-1/2\,{p_0}^{2}\eta\,{\delta}^{
2}}}=0. \label{eq6}
\end{equation}
Equation (\ref{eq6}) requires $\delta=0$, i.e., the deviation which occurs in the plane wave phase do not reside in the stripe phase.
However, as we point out in section \ref{sec4}, there is actually small deviation in the stripe phase due to the atomic interaction. As the external potential becomes more and more anisotropic, this deviation manifests itself more clearly. Deviations in anisotropic external potential is more complex for the stripe phase and we need to consider an additional degree of freedom to describe the deviation, as we will introduce in Sec. \ref{sec4}.

\section{Numerical results in the plane wave phase and explorations by manipulation of the external potential}\label{sec3}
In this section we numerically investigate the deviation pointed out in section \ref{sec2}.
As shown in the previous analytical calculations, the deviation $\delta$ is small and it is usually difficult to identify the existence of this kind of deviation by only keeping eyes on the density distribution.
In Fig. \ref{fig4}(a), we show the plane wave phase of spin-$1$ condensate. Theoretical explorations show that plane wave phase can be realized in the $^{87}$Rb condensate because the spin-exchange interaction coefficient $c_2<0$ \cite{Wang2011}.

By resorting to the density of magnetic moment along the $z$ direction:
\begin{equation}
\mathcal{F}_z=|\psi_1|^2-|\psi_{-1}|^2,
\end{equation}
we can specify this kind of deviation.
If density distribution of particles in component $\psi_1$ coincides with that in component $\psi_{-1}$, a uniform $\mathcal{F}_z$ would be expected. Otherwise, non-uniform $\mathcal{F}_z$ would appear, indicating the deviation of particle density distributions in components $\psi_1$ and $\psi_{-1}$.
We show the distribution of $\mathcal{F}_z$ in Fig. \ref{fig4}(b). From Fig. \ref{fig4}(b), it is clearly shown that $\mathcal{F}_z$ chooses positive values in the bottom right corner of the panel, i.e., the density distribution of the $\psi_{1}$ component has been dragged along direction $(\hat{x}-\hat{y})/\sqrt{2}$ ($\hat{x}$ and $\hat{y}$ are unit vectors along axis $x$ and $y$, respectively). The phase distribution in Fig. \ref{fig4}(a) shows the plane wave propagates along the direction $(-\hat{x}-\hat{y})/\sqrt{2}$. The numerical results agree with the analytical calculations in section \ref{sec2}.

In order to compare the analytical and numerical results, we start calculations from Eq. (\ref{eq2}) with $c_0=500.0$ and $c_2=-5.0$. From Fig. \ref{fig4}(a), we find the deviation of the density peaks between components with $m=1$ and $m=-1$ is $0.25$. By referring to minimum of the total energy, we can find that such a deviation corresponds to $\eta=0.006$, i.e., a width of the condensate $\Delta=3.22$, which is compatible with the size of the numerically obtained condensate. However, as the interaction grows stronger, the consistency between analytical and numerical results gets worse and worse because the gaussian configuration ansatz is no longer appropriate. In fact, the density of each component shows asymmetrical distribution referring to its peak density point.

As the small deviation can be reflected by the polarized magnetic moment distribution, possible ways to magnify the local magnetic moment would be useful. We implement an anisotropic external trapping potential, and show the response of the local magnetic moment to this anisotropism. In Fig. \ref{fig5}(a), we show the variation of the peak value of $\mathcal{F}_z$ with respect to the $\omega_\perp/\omega$, where $\omega_\perp$ is the trapping frequency in the squeezing direction. As $\omega_\perp/\omega$ increases, the local magnetic moment gets larger and larger, which is helpful to assure the small deviation in the experiments. The spin texture, i.e. half-skyrmions, is plotted in Fig. \ref{fig5}(b). For the square lattice structure of the condensate,  half-skyrmions lattices would appear, the squeezing of the external potential takes up the role of magnifying the local $z$ component of the magnetic moment, which is similar to that in the plane wave phase. We should mention that, as $\omega_\perp/\omega$ gets larger, the condensate would cross from the square lattice phase to the plane wave phase.

\section{Hidden spin textures in the stripe phase}\label{sec4}
\begin{figure}
\centering
\begin{minipage}[b]{0.5\textwidth}
\centering
\includegraphics[width=1.9in]{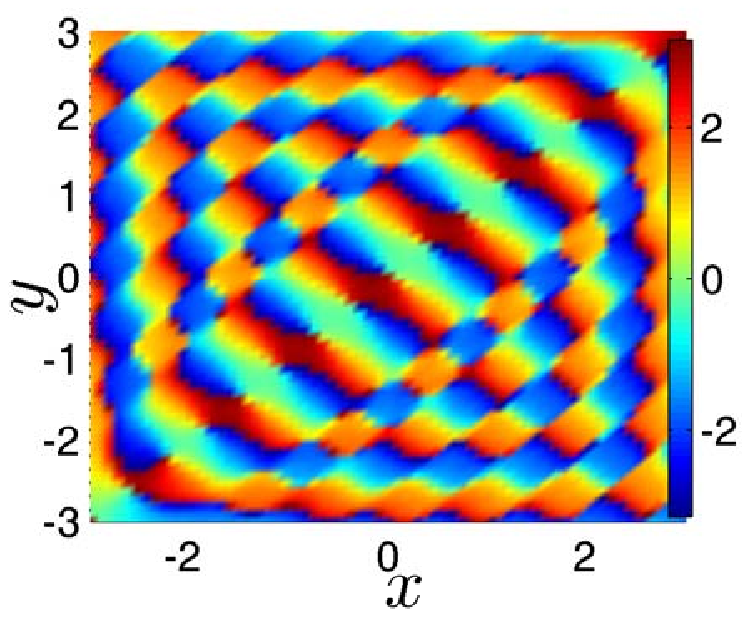}
\end{minipage}
\begin{minipage}[b]{0.5\textwidth}
\centering
\includegraphics[width=3.7in]{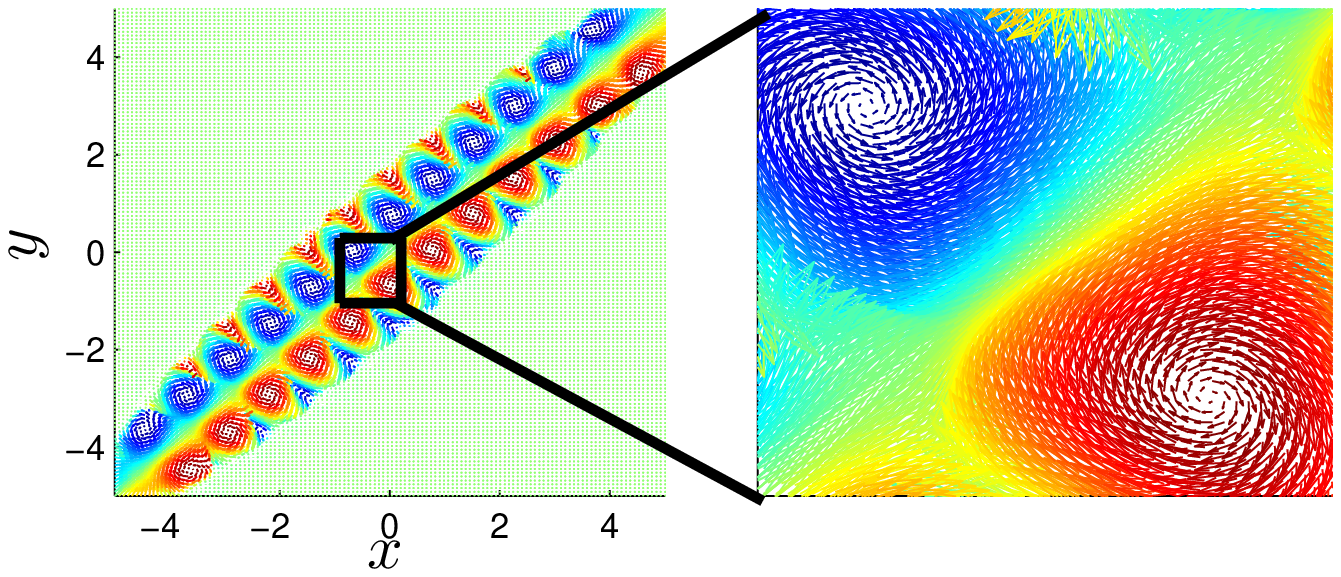}
\end{minipage}
\caption{(Color online) The panels in the bottom are the spin texture obtained by squeezing the condensate in the stripe phase with $\kappa=2.0$, $c_0=500.0$, $c_2=5.0$ and $\omega_\perp/\omega=8$, where $\omega_\perp$ is the trapping frequency in the squeezing direction. Spin texture on the right is the magnification of the squared part in the left. The panel in the top shows the corresponding phase distribution of the $\psi_1$ component in the stripe phase.}
\label{fig7}
\end{figure}
Through the analytical and numerical calculations, we know that there is deviation between the density distributions of components $\psi_1$ and $\psi_{-1}$ for the plane wave phase. This deviation occurs in the direction perpendicular to the propagating direction of the plane wave. In this section we focus on the stripe phases. The stripe phase is predicted to exist in the spinor condensate of $^{23}$Na with the spin-exchange interaction $c_2>0$ \cite{Wang2011}.

In the following, we take the parameters $\kappa=2.0$, $c_0=500.0$ and $c_2=5.0$, and the trapped stripe phase is found numerically. The density and phase distributions of component $\psi_1$ are plotted in Fig. \ref{fig6}(a). Figure \ref{fig6}(b) shows the spin texture of the stripe phase. The hidden half-skyrmions can be clearly seen in the periphery of the condensate. The spin texture of a pair of half-skyrmions is shown in the bottom of Fig. \ref{fig6}(b). By applying external manipulation on the condensate, i.e., squeezing the condensate in certain direction, the spin texture can manifest themselves more clearly, which is shown by the panels in the bottom of Fig. \ref{fig7}.
The color on the arrows represents the local magnetic moment in the $z$ direction (the modulus of the magnetic moment have been normalized to unity). As the color changes from blue to dull-red, the relative atom number changes between components $\psi_1$ and $\psi_{-1}$.  A line of paired half-skyrmions can be seen. The panel on the right is the magnification of the squared area in the panel on the left. From the spin texture in Fig. \ref{fig7}, there seems to be small deviation between components $\psi_1$ and $\psi_{-1}$ because of the evident color contrast in the paired half-skyrmions line.

In order to understand the numerical results in Fig. \ref{fig7} qualitatively, we introduce the following ansatz:
\begin{sequation}
\begin{split}
\psi_{1}\!&=\frac{1}{2}\!C\!\left(\!-{e}^{i\!p_0\! x\!}{e}^{-{p_0}^{2}\!\eta\!
\left(\!{x}^{2}\!+R\left(\!y\!-y_0\!+\delta\!\right)^{2}\!\right)\!}
+{e}^{-i\!p_0\! x\!}\!{e}^{-{p_0}^{2}\!\eta\!\left(\!{x}^{2}\!+
\!R\left(\!y\!-y_0\!-\!\delta\!\right)^{2}\!\right)\!}\right),\!\\
\psi_{0}&=\frac{\sqrt{2}}{2}\!C\!{{e}^{-\!{p_0\!}^{2}\!\eta\!\left(\!{x}^{2}\!+R{y}^{2}\!
 \right)\!}}\left(\!{{e}^{i\!p_0\! x\!}}\!+{{e}^{-i\!p_0\! x\!}}
 \right),\!\\
\psi_{-1}&=\frac{1}{2}C\!\left(\!-{{e}^{i\!p_0\! x\!}}{{e}^{-\!{p_0\!}^{2}\eta\!
 \left(\!{x}^{2}\!+ R\left(\! {y_0}\!+y\!-\delta\!\right)\!^{2}\! \right)\!}}+\!{
{e}^{-i\!p_0\!\,x\!}}{{e}^{-\!{p_0\!}^{2}\!\eta\!\,\! \left(\!{x}^{2}\!+
 R\left(\! {y_0}\!+y\!+\delta\!\right)\! ^{2}\right)\!}}\right),\! \label{eq10}
\end{split}
\end{sequation}
where, $C=p_0\,\sqrt {{\frac {\eta}{\pi }}}$ is the normalization coefficient, and $R$ can fix the ratio of the size scales of the condensate in $x$ and $y$ directions. Repeating the same calculations as in Sec. \ref{sec2}, we can obtain the dependence of the energy on parameters $\delta$ and $y_0$. Here $\delta$ represents separation between the two gaussian packets bearing counter-propagating plane waves and $y_0$ characterizes the average separation of packets in $m=1$ and $m=-1$ components.

\begin{figure}
\centering
\subfigure{(a)}
\begin{minipage}[b]{0.5\textwidth}
\centering
\includegraphics[width=2.2in]{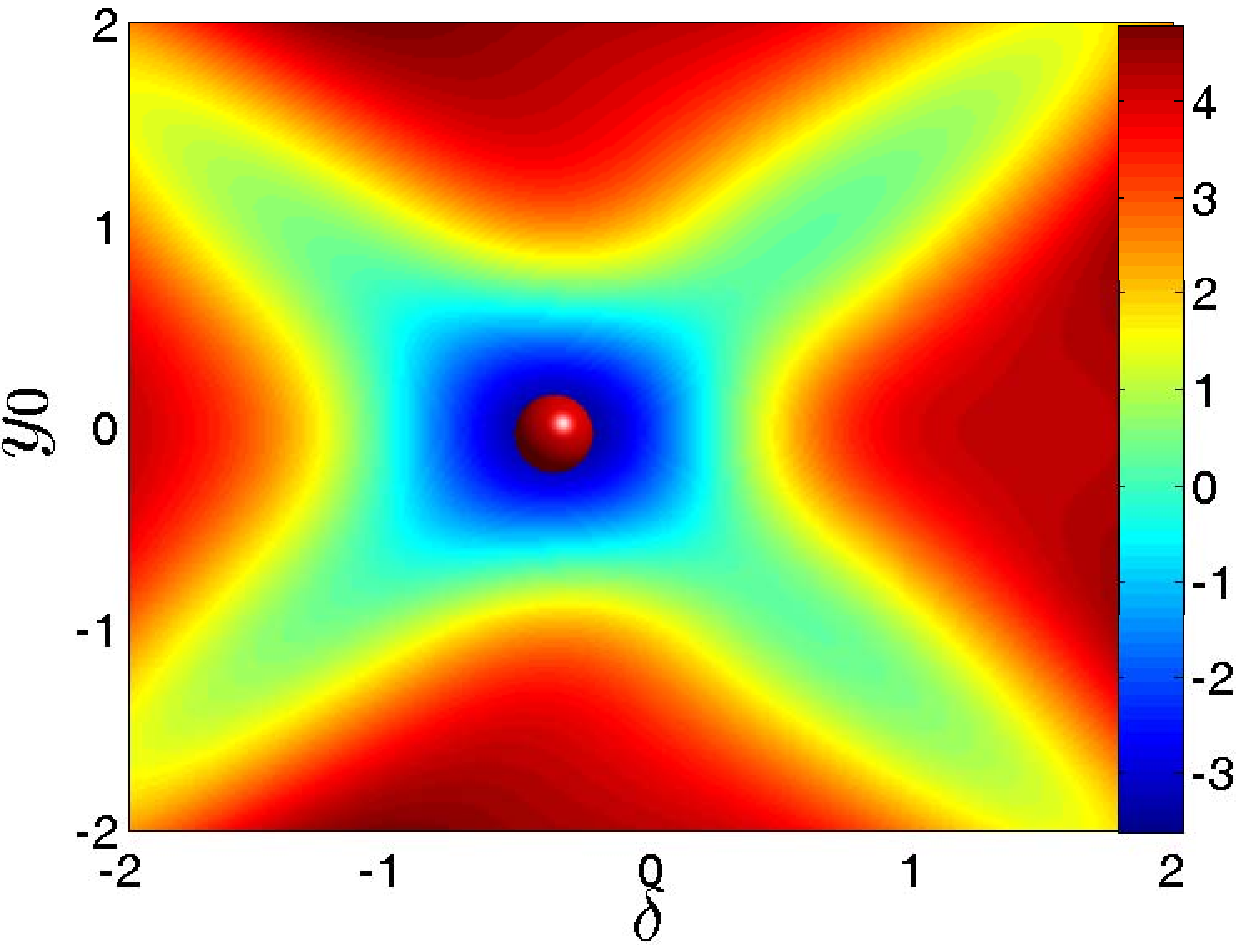}
\end{minipage}
\subfigure{(b)}
\begin{minipage}[b]{0.5\textwidth}
\centering
\includegraphics[width=2.5in]{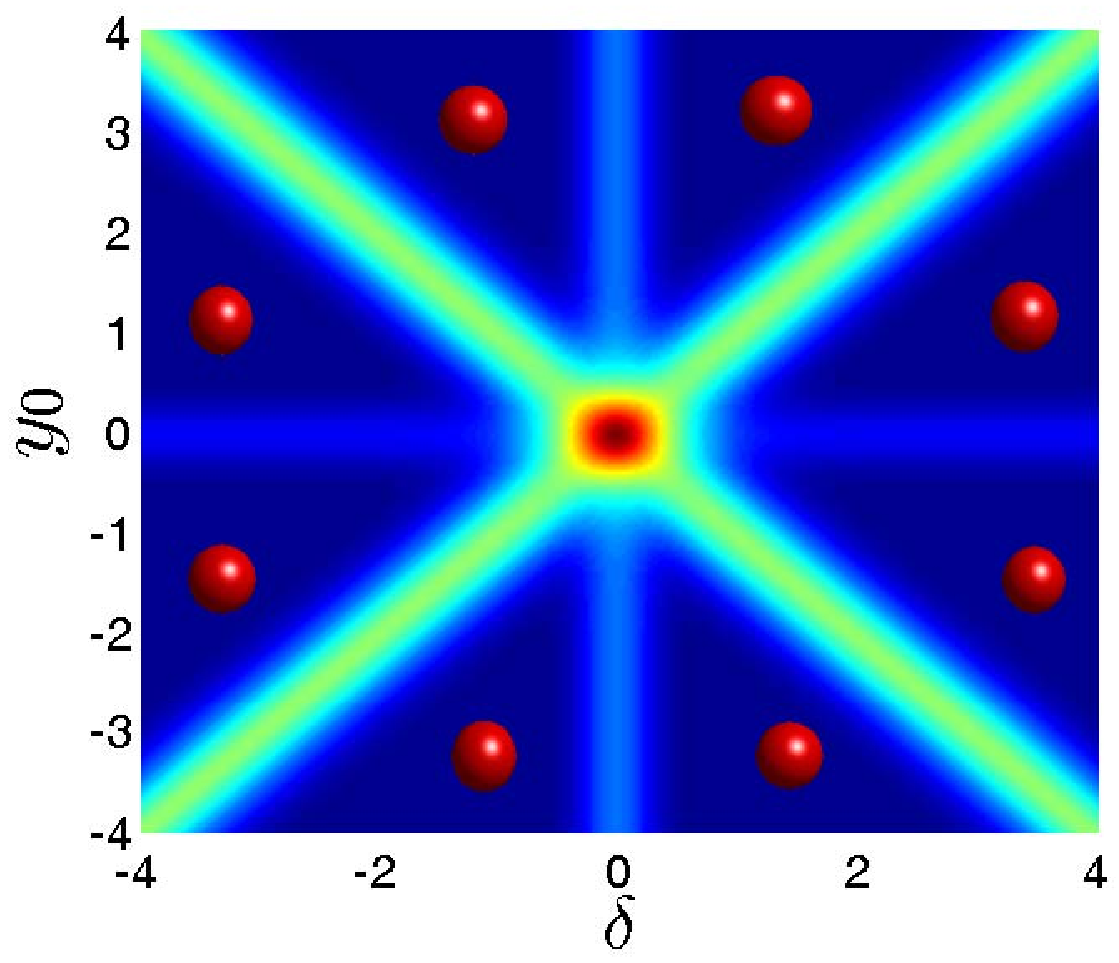}
\end{minipage}
\caption{(Color online) The dependence of (a) the $H_{trap}+H_s$ and (b) $H_{int}$ on $\delta$ and $y_0$, with parameters $\kappa=2.0$, $\eta=0.0096$, $c_0=500.0$, $c_2=5.0$, and R=64.}
\label{fig8}
\end{figure}

From Fig. \ref{fig8}(a), where the parabolic potential and SO parts of the energy is plotted as a function of $\delta$ and $y_0$, we can see that the energetic minimum is shifted in the negative direction of the $\delta$ axis (marked by light-red ball), while there is no shift in the $y_0$ axis. By counting in the non-zero $\delta$ in the ansatz (\ref{eq10}), the phase distribution shows the same configuration with that of the numerical result in Fig. \ref{fig7}. In Ref. \cite{Tomoki}, a variational ansatz representing a modified striped state was proposed. The phase distribution in the present paper is consistent with the particle current found in Ref. \cite{Tomoki}.

Thus far, the roles of the interaction have not been included. In Fig. \ref{fig8}(b), we plot the interaction energy $E_{int}$ as a function of $\delta$ and $y_0$. There are eight energetic minimums (marked by light-red ball in the quasi triangular contour) around the center. Therefore, the role of the interaction is to drag the energetic minimum off the center ($\delta=y_0=0$). The color distribution (blue to dull-red) in Fig. \ref{fig7}(b), which represents non-uniform magnetic momentum component $\mathcal{F}_z$, indicates a non-zero $y_0$ has been chosen.

In the end, we discuss and exhibit some key experimental parameters. The real SO coupling strength in the present paper is $\gamma=\kappa\sqrt{\hbar m \omega}$. A Rashba SO coupling can be induced
in spinor $^{87}$Rb cold atoms \cite{LinNature}. Experimentally, a tight harmonic confinement along $z$ is necessary to realize the pancake-like condensate. If a spinor condensate composed of $N=10^4$ atoms is realized and the radial and axial trapping frequencies are chosen as $2\pi\times 850Hz$ and $2\pi\times 20Hz$, respectively, the corresponding s-scattering lengths are $a_0=68.76a_B$ (total spin channel $F_{total}=0$) and $a_2=70.87a_B$ (total spin channel $F_{total}=2$) for $^{87}$Rb cold atoms and $a_0=138.28a_B$ and $a_2=134.21a_B$ for $^{23}$Na cold atoms.

\section{CONCLUSIONS} \label{sec5}
In summary, we have analytically and numerically studied deviations between the density distributions of the $\psi_1$ and $\psi_{-1}$ components in SO coupled spin-$1$ BEC with external potential, which results in the hidden spin textures in the trapped plane wave and stripe phases. The polarized magnetic moment distribution in $z$ direction, $\mathcal{F}_z$, can mark the existence of the deviation. Manipulation of the external potential can be utilized to magnify the local $\mathcal{F}_z$ in both the plane wave and the square lattice phases. Although $\mathcal{F}_z$ in the stripe phase is small, nontrivial spin texture dwells in the condensate. As the squeezing of the external trap gets harder and harder, the hidden half-skyrmions align themselves, forming into a paired half-skyrmions line. With the mean field energy minimized, the deviation that we obtained is the consequence of the cooperation of all the components of the condensate in the presence of the SO coupling and the external trap.
Thus, for spinor condensate with spin-orbit coupling strength $\kappa\sim1$ and comparable external traps, the usual plane wave phase or stripe phase are revised, bringing nontrivial detail structures. The present paper should help us to understand the cooperative roles of the SO coupling and the external potential in determining the ground state of the spinor condensates.

\textbf{Acknowledgments}
This work was supported by the NKBRSFC under Grants No. 2011CB921502, No. 2012CB821305, No. 2009CB930701, No. 2010CB922904, the NSFC under Grants No. 10934010, No. 60978019, and the NSFC-RGC under Grants No. 11061160490 and No. 1386-N-HKU748/10.  H.W. was supported by the NSFC under Grant No. 10901134. Ji was supported by the NCET and NSFC under Grant No. 11704175 and No. 10904096.

\end{document}